\documentstyle[12pt]{article}
\def\photonatomright{\begin{picture}(3,1.5)(0,0)
                                \put(0,-0.75){\tencircw \symbol{2}}
                                \put(1.5,-0.75){\tencircw \symbol{1}}
                                \put(1.5,0.75){\tencircw \symbol{3}}
                                \put(3,0.75){\tencircw \symbol{0}}
                      \end{picture}
                     }

\def\photonright{\begin{picture}(30,1.5)(0,0)
                     \multiput(0,0)(3,0){10}{\photonatomright}
                  \end{picture}
                 }

\def\photonrighthalf{\begin{picture}(30,1.5)(0,0)
                     \multiput(0,0)(3,0){5}{\photonatomright}
                  \end{picture}
                 }

\def\fermionuphalf{\begin{picture}(1,15)(0,0)
                         \put(0,0){\vector(0,1){7.5}}
                         \put(0,7.5){\line(0,1){7.5}}
                   \end{picture}
                  }

\def\fermiondown{\begin{picture}(1,30)(0,-30)
                       \put(0,0){\vector(0,-1){15}}
                       \put(0,-15){\line(0,-1){15}}
                 \end{picture}
                }

\def\fermionull{\begin{picture}(30,15)(0,0)
                        \put(0,0){\vector(-2,1){15}}
                        \put(-15,7.5){\line(-2,1){15}}
                  \end{picture}
                 }
\def\fermionullhalf{\begin{picture}(15,7.5)(0,0)
                        \put(0,0){\vector(-2,1){7.5}}
                        \put(-7.5,3.75){\line(-2,1){7.5}}
                  \end{picture}
                 }
\def\fermionurr{\begin{picture}(30,15)(0,0)
                        \put(-30,-15){\vector(2,1){15}}
                        \put(-15,-7.5){\line(2,1){15}}
                  \end{picture}
                 }
\def\fermionurrhalf{\begin{picture}(15,7.5)(0,0)
                        \put(-15,-7.5){\vector(2,1){7.5}}
                        \put(-7.5,-3.75){\line(2,1){7.5}}
                  \end{picture}
                 }
\def\fermiondrr{\begin{picture}(30,15)(0,0)
                        \put(0,0){\vector(2,-1){15}}
                        \put(15,-7.5){\line(2,-1){15}}
                  \end{picture}
                 }
\def\fermiondrrhalf{\begin{picture}(15,7.5)(0,0)
                        \put(0,0){\vector(2,-1){7.5}}
                        \put(7.5,-3.75){\line(2,-1){7.5}}
                  \end{picture}
                 }
\def\fermiondll{\begin{picture}(30,15)(0,0)
                        \put(30,15){\vector(-2,-1){15}}
                        \put(15,7.5){\line(-2,-1){15}}
                  \end{picture}
                 }
\def\fermiondllhalf{\begin{picture}(15,7.5)(0,0)
                        \put(15,7.5){\vector(-2,-1){7.5}}
                        \put(7.5,3.75){\line(-2,-1){7.5}}
                  \end{picture}
                 }

\newenvironment{Feynman}[3]{\begin{center}
                            \setlength{\unitlength}{#3 mm}
                            \begin{picture}(#1)(#2)
                            \thicklines
                           }{\end{picture} \end{center}}

\catcode`\@=11
\long\def\@makefntext#1{
\protect\noindent \hbox to 3.2pt {\hskip-.9pt
$^{{\ninerm\@thefnmark}}$\hfil}#1\hfill}		

\def\@makefnmark{\hbox to 0pt{$^{\@thefnmark}$\hss}}  

\def\ps@myheadings{\let\@mkboth\@gobbletwo
\def\@oddhead{\hbox{}
\rightmark\hfil\ninerm\thepage}
\def\@oddfoot{}\def\@evenhead{\ninerm\thepage\hfil
\leftmark\hbox{}}\def\@evenfoot{}
\def\sectionmark##1{}\def\subsectionmark##1{}}

\renewcommand{\thefootnote}{\fnsymbol{footnote}}

\newcounter{sectionc}\newcounter{subsectionc}\newcounter{subsubsectionc}
\renewcommand{\section}[1] {\vspace*{0.6cm}\addtocounter{sectionc}{1}
\setcounter{subsectionc}{0}\setcounter{subsubsectionc}{0}\noindent
	{\normalsize\bf\thesectionc. #1}\par\vspace*{0.4cm}}
\renewcommand{\subsection}[1] {\vspace*{0.6cm}\addtocounter{subsectionc}{1}
	\setcounter{subsubsectionc}{0}\noindent
	{\normalsize\it\thesectionc.\thesubsectionc. #1}\par\vspace*{0.4cm}}
\renewcommand{\subsubsection}[1]
{\vspace*{0.6cm}\addtocounter{subsubsectionc}{1}
\noindent {\normalsize\rm\thesectionc.\thesubsectionc.\thesubsubsectionc.
	#1}\par\vspace*{0.4cm}}

\newcounter{appendixc}
\newcounter{subappendixc}[appendixc]
\newcounter{subsubappendixc}[subappendixc]

\renewcommand{\appendix}[1] {\vspace*{0.6cm}
        \refstepcounter{appendixc}
        \setcounter{figure}{0}
        \setcounter{table}{0}
        \setcounter{equation}{0}
        \renewcommand{\thefigure}{\Alph{appendixc}.\arabic{figure}}
        \renewcommand{\thetable}{\Alph{appendixc}.\arabic{table}}
        \renewcommand{\theappendixc}{\Alph{appendixc}}
        \renewcommand{\theequation}{\Alph{appendixc}.\arabic{equation}}
        \noindent{\bf Appendix \theappendixc #1}\par\vspace*{0.4cm}}

\def\abstracts#1{{
\centering{\begin{minipage}{12.2truecm}\footnotesize\baselineskip=12pt\noindent
	\centerline{\footnotesize ABSTRACT}\vspace*{0.3cm}
	\parindent=0pt #1
	\end{minipage}}\par}}

\renewenvironment{thebibliography}[1]
	{\begin{list}{\arabic{enumi}.}
	{\usecounter{enumi}\setlength{\parsep}{0pt}
\setlength{\leftmargin 1.25cm}{\rightmargin 0pt}
	 \setlength{\itemsep}{0pt} \settowidth
	{\labelwidth}{#1.}\sloppy}}{\end{list}}

\newcounter{itemlistc}
\newcounter{romanlistc}
\newcounter{alphlistc}
\newcounter{arabiclistc}

\newcommand{\fcaption}[1]{
        \refstepcounter{figure}
        \setbox\@tempboxa = \hbox{\footnotesize Fig.~\thefigure. #1}
        \ifdim \wd\@tempboxa > 6in
           {\begin{center}
        \parbox{6in}{\footnotesize\baselineskip=12pt Fig.~\thefigure. #1}
            \end{center}}
        \else
             {\begin{center}
             {\footnotesize Fig.~\thefigure. #1}
              \end{center}}
        \fi}

\newcommand{\tcaption}[1]{
        \refstepcounter{table}
        \setbox\@tempboxa = \hbox{\footnotesize Table~\thetable. #1}
        \ifdim \wd\@tempboxa > 6in
           {\begin{center}
        \parbox{6in}{\footnotesize\baselineskip=12pt Table~\thetable. #1}
            \end{center}}
        \else
             {\begin{center}
             {\footnotesize Table~\thetable. #1}
              \end{center}}
        \fi}

\def\@citex[#1]#2{\if@filesw\immediate\write\@auxout
	{\string\citation{#2}}\fi
\def\@citea{}\@cite{\@for\@citeb:=#2\do
	{\@citea\def\@citea{,}\@ifundefined
	{b@\@citeb}{{\bf ?}\@warning
	{Citation `\@citeb' on page \thepage \space undefined}}
	{\csname b@\@citeb\endcsname}}}{#1}}

\newif\if@cghi
\def\cite{\@cghitrue\@ifnextchar [{\@tempswatrue
	\@citex}{\@tempswafalse\@citex[]}}
\def\citelow{\@cghifalse\@ifnextchar [{\@tempswatrue
	\@citex}{\@tempswafalse\@citex[]}}
\def\@cite#1#2{{$\null^{#1}$\if@tempswa\typeout
	{IJCGA warning: optional citation argument
	ignored: `#2'} \fi}}

 1
 1
 1

\font\ninerm=cmr9

\newcommand{\nll}{\nonumber \\}

\newcommand{\bq}{\begin{equation}}
\newcommand{\eq}{\end{equation}}
\newcommand{\ba}{\begin{eqnarray}}
\newcommand{\ea}{\end{eqnarray}}

\newcommand{\nobody}{\rule{0ex}{1ex}}
\newcommand{\nobodyfrac}{\frac{\nobody}{\nobody}}

\begin{document}
\begin{flushright}
LMU-11/95\\
hep-ph/9504358
\vspace{2cm}\\
\end{flushright}
\centerline{\normalsize\bf SEMIANALYTIC DISTRIBUTIONS}
\baselineskip=16pt
\centerline{\normalsize\bf IN FOUR FERMION NEUTRAL CURRENT PROCESSES
\footnote{To appear in the proceedings of Ringberg workshop
``Perspectives for electroweak interactions in $e^+e^-$ collisions'',
Ringberg, Germany, February 1995.}
}

\centerline{\footnotesize Arnd Leike
\footnote{Supported by the German Federal Ministry for Research
and Technology under contract No.~05~6MU93P}
}
\baselineskip=13pt
\centerline{\footnotesize\it Ludwig-Maximilians-Universit\"at,
Theresienstr. 37, D--80333 M\"unchen, Germany
}
\baselineskip=12pt
\centerline{\footnotesize E-mail: leike@cernvm.cern.ch}

\vspace*{0.9cm}
\abstracts{
Analytical formulae for triple differential distributions
${\rm d}^3\sigma/({\rm d}\cos\theta{\rm d}s_1{\rm d}s_2)$ in
the neutral current
process $e^+e^-\rightarrow f_1\bar f_1 f_2\bar f_2$ are given.
They allow to obtain angular distributions, rapidity distributions and
transversal momentum distributions of one fermion pair
by only two numerical integrations.
Cuts can be applied to the integration variables. }
\normalsize\baselineskip=15pt
\setcounter{footnote}{0}
\renewcommand{\thefootnote}{\alph{footnote}}

\section{Introduction}
Many events of four fermion final states are expected to be observed at LEPII
and  NLC (next linear collider). The accuracy of the theoretical description
of such processes should  match the future experimental precision.
As is known from LEPI, both, semianalytical programs and
Monte Carlo generators are needed to understand the physics of experimental
data.

Semianalytical calculations lead to fast {\tt FORTRAN} codes with high
numerical precision.
Although a semianalytical calculation will never be able to simulate a
real detector, it should be as differential as reasonable. This allows to
predict observables closer to the experiment and to make more meaningful
comparisons with other groups \cite{bkp,pittau}.

The semianalytic approach
has been applied earlier to
 the calculation of four fermion final states
of neutral current process \cite{blr}\
$e^+e^-\rightarrow f_1\bar f_1f_2\bar f_2$  and
Higgs production in the same reaction \cite{higgs}.
There, analytical formulae for the {\it double} differential cross section
were given. The total cross section was obtained by two
numerical integrations.
Differential distributions over the invariant energy of one fermion pair
($f_1,\bar f_1$) were obtained by one integration.

In this contribution, we present analytical formulae for {\it triple}
differential cross sections. In addition to the invariant energies of
the two fermion pairs, we keep the cosine of the angle between the 3-vector
of the momentum of one fermion pair and the beam axis as an additional
parameter. The resulting analytical formulae
turn out to be not much longer than the double differential ones.
However, the formulae presented here
 may be used to get new distributions by two numerical integrations:
angular distributions , rapidity
distributions and transversal momentum distributions
of one fermion pair. These distributions
 can be obtained including additional cuts on the remaining integration
variables.

We start with the process-independent description of the phase space in
section~2 and give formulae for the cross section
and distributions in section~3.
Section~4 contains a brief summary.
Detailed formulae for the kinematical functions of the triple differential
distributions are given in the appendix.
Numerical applications will be presented elsewhere.

%
\section{Phase space}
We are interested in the cross section and in distributions of the reaction
\bq
\label{reaction}
e^+(k_1)e^-(k_2)\rightarrow f_1(p_1)\bar f_1(p_2) f_2(p_3)\bar f_2(p_4)
\eq
with $p_i^2=m_i^2,\ \ i=1,\dots ,4$.

The eightdimensional phase space of the four particle final
state is parametrized as
%
\begin{eqnarray}
\label{domega}
d\Gamma
&=& \prod_{i=1}^4\frac{d^3p_i}{2p_{i}^0}
 \times  \delta^4(k_1+k_2-\sum_{i=1}^4 p_i)
\nonumber
\\
&=&~2\pi\frac{\sqrt{\lambda(s,s_1,s_2)}}{8s}
\frac{\sqrt{\lambda(s_1,m_1^2,m_2^2)}}{8s_1}
\frac{\sqrt{\lambda(s_2,m_3^2,m_4^2)}}{8s_2}
{\rm d} s_1 {\rm d} s_2 {\rm d} \Omega {\rm d} \Omega_1 {\rm d} \Omega_2.
\end{eqnarray}
$\lambda$ is the kinematical function,
\ba
\lambda(a,b,c) = a^2+b^2+c^2-2ab-2ac-2bc,\ \ \
\lambda \equiv \lambda(s,s_1,s_2),
\label{lambda}
\ea
and the invariants $s,\ s_1$ and $s_2$ are
\bq
s=(k_1+k_2)^2,\ \
s_1=(p_1+p_2)^2,\ \
 s_2=(p_3+p_4)^2.
\eq
We integrate over every spherical angle $\Omega$ ($\Omega_1,\ \Omega_2$)
of the $e^+e^-$ pair ($f_1\bar f_1,\ f_2\bar f_2$ pair)
 in its rest frame.
The spherical angles are decomposed in two integrations,
${\rm d}\Omega_i={\rm d}\cos\theta_i{\rm d}\phi_i$ with the kinematical
ranges
\bq
-1 \le \cos\theta_i \le 1,\ \ \  0 \le \phi_i \le 2\pi.
\eq
In particular, we have
${\rm d}\Omega={\rm d}\cos\theta{\rm d}\phi$, where
$\cos\theta=c$ is the cosine of the angle between the 3-vectors
($\vec{p}_1+\vec{p}_2$) and $\vec{k}_1$.
In the case without transversal beam polarization, the integration over
the rotation angle $\phi$ around the beam axis is trivial, giving $2\pi$.

In the following, we will integrate the squared matrix element over the
four remaining angular variables
$\cos\theta_1,\phi_1,\cos\theta_2$ and $\phi_2$. This results
to analytic functions depending on the remaining variables
 $c,\ s_1$ and $s_2$.
They enable us to calculate not only total cross sections but also
angular distributions ${\rm d}\sigma/{\rm d}c$,
transversal momentum distributions
${\rm d}\sigma/{\rm d}p_{iT}$
or rapidity distributions ${\rm d}\sigma/{\rm d}y_i$
of the ``compound particle''
($f_i,\bar f_i,\ i=1,2$).
These new distributions are useful for comparisons with Monte Carlo
programs and for more reliable experimental predictions.

The transformation of the phase space volume to the new integration
variables is
\ba
\int_{-1}^{1}{\rm d}c\int_{s_1^-}^{s_1^+}{\rm d}s_1
\int_{s_{2}^-(s_1)}^{s_{2}^+(s_1)}{\rm d}s_{2}
&=&
2\sqrt{s}\int_{y_1^-}^{y_1^+}{\rm d}y
\int_{p_{1T}^-(y_1)}^{p_{1T}^+(y_1)}{\rm d}p_{1T}
\int_{s_{2}^-(p_{1T},y_1)}^{s_{2}^+(p_{1T},y_1)}{\rm d}s_{2}
\nll
&=&
2\sqrt{s}\int_{p_{1T}^-}^{p_{1T}^+}{\rm d}p_{1T}
\int_{y_{1}^-(p{_1T})}^{y_{1}^+(p_{1T})}{\rm d}y_{1}
\int_{s_{2}^-(p_{1T},y_1)}^{s_{2}^+(p_{1T},y_1)}{\rm d}s_{2}
\ea
with similar formulae for $p_{2T}$ and $y_2$.
The old variables $c$ and $s_1$ have to be expressed through the new
variables $p_{1T}$ and $y$:
\ba
\label{vartrafo}
c&=&\sqrt{\frac{1+p_{1T}^2/(m_1+m_2)^2}{1+\coth^2(y)\, p_{1T}^2/(m_1+m_2)^2}},
\nll
s_1&=&2\sqrt{s}\sqrt{(m_1+m_2)^2+p_{1T}^2}\, \cosh y +s_2-s
\ea
The kinematical ranges of the integration variables $s_1$ and $s_2(s_1)$ are
\ba\begin{array}{rclrcl}
s_1^- &=& (m_1+m_2)^2, &     s_1^+ &=& (\sqrt{s}-m_3-m_4)^2,\\
s_2^-(s_1) &=& (m_3+m_4)^2, & s_2^+(s_1) &=& (\sqrt{s}-\sqrt{s_1})^2.
\end{array}
\ea
For simplicity, we give the other kinematical borders only  for the case
of massless final fermions:
\ba
\begin{array}{rclrcl}
y_1^-&=&-\infty, & y_1^+&=&\infty, \\
p_{1T}^-(y_1)&=&0, & p_{1T}^+(y_1)&=&\sqrt{s}/\cosh y,
\\
p_{1T}^-&=&0, & p_{1T}^+&=&\sqrt{s}, \\
y_1^-(p_{1T})&=&-{\rm arcosh}\frac{\sqrt{s}}{p_{1T}}, &
y_1^+(p_{1T})&=&{\rm arcosh}\frac{\sqrt{s}}{p_{1T}}
\end{array}
\ea
The borders of $s_2(y_1,p_{1T})$ are
\bq
\label{borders2}
s_2^-(y_1,p_{1T}) =
{\rm max}\left\{\nobodyfrac 0,s-2\sqrt{s}\, p_{1T}\cosh y\right\},
\ \ \
s_2^+(y_1,p_{1T}) = (\sqrt{s}-p_{1T}\cosh y)^2.
\eq

%
\begin{figure}[tbh]
\begin{Feynman}{180,70}{0,0}{0.8}
%
\put(15,7.5){\fermiondllhalf}
\put(15,52.5){\fermiondrrhalf}
\put(30,15){\fermiondown}
\put(30,45){\photonright}
\put(30,15){\photonright}
\put(30,45){\circle*{1.5}}
\put(30,15){\circle*{1.5}}
\put(60,15){\circle*{1.5}}
\put(60,45){\circle*{1.5}}
\put(75,7.5){\fermionullhalf}
\put(75,22.5){\fermionurrhalf}
\put(75,37.5){\fermionullhalf}
\put(75,52.5){\fermionurrhalf}
\put(32,28){$e$}
\put(40,49){$\gamma,\ Z$}
\put(40,09){$ \gamma, \ Z$}
\put(77,51){$ f_1$}
\put(77,35){${\bar f_1}$}
\put(77,21){${ f_2}$}
\put(77, 5){${\bar f_2}$}
%
\put(105,7.5){\fermiondllhalf}
\put(105,52.5){\fermiondrrhalf}
\put(120,15){\fermiondown}
\put(120,45){\photonright}
\put(120,15){\photonright}
\put(120,45){\circle*{1.5}}
\put(120,15){\circle*{1.5}}
\put(150,15){\circle*{1.5}}
\put(150,45){\circle*{1.5}}
\put(165,7.5){\fermionullhalf}
\put(165,22.5){\fermionurrhalf}
\put(165,37.5){\fermionullhalf}
\put(165,52.5){\fermionurrhalf}
\put(122,28){$e$}
\put(130,49){$\gamma,\ Z$}
\put(130,09){$\gamma,\ Z$}
\put(167,51){$ f_2$}
\put(167,35){${\bar f_2}$}
\put(167,21){${ f_1}$}
\put(167, 5){${\bar f_1}$}
\end{Feynman}
\vspace*{.2cm}
{\it
{\bf Fig.~1:} The {\tt crab} diagrams.
}
\end{figure}

%
\section{Cross sections and distributions}
We present the formulae for the triple differential distribution
of the reaction (\ref{reaction}),
where only neutral gauge bosons are exchanged.
The initial electron-positron pair and the final fermions
$f_1$ and $f_2$ are all different and none of them belongs to
one electroweak multiplet. The reaction (\ref{reaction}) is then described
by 24 Feynman diagrams which exchange photons and $Z$-bosons.
In the case of four quarks in the final state, there are 8 additional diagrams
with gluon exchange. One can divide these diagrams in three different types.
The first type which we call {\tt crab} (these diagrams
are also called conversion
diagrams) is shown in fig.~1. These diagrams can
be
considered as $e^+e^-\rightarrow f_1\bar f_1$ with initial state radiation of
a $f_2\bar f_2$-pair (or $e^+e^-\rightarrow f_2\bar f_2$ with initial state
radiation of a $f_1\bar f_1$-pair). The second type which we call {\tt deers}
(they are also called annihilation diagrams)
can be considered as $e^+e^-\rightarrow f_1\bar f_1$ with radiation of
a $f_2\bar f_2$-pair from the final state ({\tt $f_1$-deer}), see fig.~2.
A different
set of diagrams is obtained for $e^+e^-\rightarrow f_2\bar f_2$ with
radiation of a $f_1\bar f_1$-pair from the final state ({\tt $f_2$-deer}).
The process is symmetric in $f_1$ and $f_2$.

\begin{figure}[bhtp]
\begin{Feynman}{180,60}{-5.0,0}{0.8}
%
\put(-10,45){\fermiondrr}
\put(-10,15){\fermiondll}
\put(20,30){\photonrighthalf}
\put(65,15){\fermionull}
\put(20,30){\circle*{1.5}}
\put(35,30){\circle*{1.5}}
\put(35,45){\circle*{1.5}}
\put(50,45){\circle*{1.5}}
\put(35,30){\fermionuphalf}
\put(65,60){\fermionurr}
\put(35,45){\photonrighthalf}
\put(65,52.5){\fermionurrhalf}
\put(65,37.5){\fermionullhalf}
\put(22,24.5){$\gamma,\ Z$}
\put(38,39){$\gamma,\ Z$}
\put(67,59){$ f_1$}
\put(67,51){$ f_2$}
\put(67,35){${\bar f_2}$}
\put(67,13){${\bar f_1}$}
%
\put(90,45){\fermiondrr}
\put(90,15){\fermiondll}
\put(120,30){\photonrighthalf}
\put(120,30){\circle*{1.5}}
\put(135,30){\circle*{1.5}}
\put(135,15){\circle*{1.5}}
\put(150,15){\circle*{1.5}}
\put(135,15){\fermionuphalf}
\put(165,00){\fermionull}
\put(165,45){\fermionurr}
\put(135,15){\photonrighthalf}
\put(165,22.5){\fermionurrhalf}
\put(165,07.5){\fermionullhalf}
\put(122,24.5){$\gamma,\ Z$}
\put(138,19){$\gamma,\ Z$}
\put(167,44){$ f_1$}
\put(167,21){$ f_2$}
\put(167, 5){${\bar f_2}$}
\put(167,-2){${\bar f_1}$}
\end{Feynman}
\vspace*{.2cm}
{\it {\bf Fig.~2:}
The {\tt $f_1$-deer} diagrams. The {\tt $f_2$-deers} may be obtained by
interchanging $f_1$ and $f_2$.
}
\end{figure}

%
The squared matrix element of the considered reaction is calculated using
the symbolic manipulation program {\tt FORM}~\cite{form}.
As a result, analytical formulae for the triple differential cross section
are obtained,
\ba
\frac{{\rm d}^3\sigma(c,s,s_1,s_2)}{{\rm d}c{\rm d}s_1{\rm d}s_2}
 =
\sum_{k=1}^6
\frac{{\rm d}^3\sigma_k(c,s,s_1,s_2)}{{\rm d}c{\rm d}s_1{\rm d}s_2}.
\label{tsig}
\ea
%
The summation runs over the squares and interferences between the three
different types of diagrams:
$k=1$ corresponds to the square of {\tt crabs},
and $k=2,3$
to that of {\tt $f_1$-deers} and {\tt $f_2$-deers}.
Further, $k=4,5$ and 6 correspond to the interferences between
{\tt $f_1$-deers} and {\tt $f_2$-deers}, between
{\tt{crabs}} and {\tt $f_2$-deers}, and between
{\tt{crabs}} and {\tt $f_1$-deers}, respectively.

For the squares of diagrams of one group ($k=1,2,3$), we obtain
\ba
\frac{{\rm d}^3\sigma_1(c,s,s_1,s_2)}{{\rm d}c{\rm d}s_1{\rm d}s_2} &=&
\frac{\sqrt{\lambda}}{\pi s^2}
 C_{422}(e,s;f_1,s_1;f_2,s_2){\cal T}_1 {\cal G}_{422}^c(c;s;s_1,s_2)~,\nll
\frac{{\rm d}^3\sigma_2(c,s,s_1,s_2)}{{\rm d}c{\rm d}s_1{\rm d}s_2} &=&
\frac{\sqrt{\lambda}}{\pi s^2}
 C_{422}(f_1,s_1;f_2,s_2;e,s){\cal T}_2 {\cal G}_{422}^c(c;s_1;s_2,s)~,\nll
\frac{{\rm d}^3\sigma_3(c,s,s_1,s_2)}{{\rm d}c{\rm d}s_1{\rm d}s_2} &=&
\frac{\sqrt{\lambda}}{\pi s^2}
 C_{422}(f_2,s_2;e,s;f_1,s_1){\cal T}_3 {\cal G}_{422}^c(c;s_2;s,s_1)~.
\ea
The color factor is ${\cal T}_k= N_c(f_1) N_c(f_2)$ for all $k=1,\dots,6$, if
photons or $Z$-bosons are exchanged.
$N_c(f)$ is equal to unity for leptons and three for quarks.
In the case of two different quarks
in the final state, also gluon exchange is possible.
The interference between
diagrams with one gluon and those without a gluon gives ${\cal T}_k= 0$, the
interference between two diagrams with a gluon gives
${\cal T}_k= 2/9,\ k=2,3,4$. The corresponding contributions have to be
added to $\sigma_2,\ \sigma_3$ and $\sigma_4$.
The kinematical function ${\cal G}_{422}^c(c;s;s_1,s_2)$
is given in the Appendix.
The function $C_{422}$ contains couplings and gauge boson propagators,
\ba
\label{c422}
C_{422}(e,s;f_1,s_1;f_2,s_2) &=& \frac{2 s_1 s_2}{(6\pi^2)^2} \;
  \Re e   \sum_{V_i,V_j,V_k,V_l=\gamma,Z}
\frac{1}{D_{V_i}(s_1)}      \frac{1}{D_{V_j}(s_2)}
\frac{1}{D_{V_k}^*(s_1)}    \frac{1}{D_{V_l}^*(s_2)}
\nll & &
\times~
\left[ L(e,V_i)L(e,V_k)L(e,V_j)L(e,V_l)
\right.\nll & &\hspace{2cm}\left.
+R(e,V_i)R(e,V_k)R(e,V_j)R(e,V_l)\right]
\nll & &
\times~
\left[ L(f_1,V_i)L(f_1,V_k)+R(f_1,V_i)R(f_1,V_k)\right]
\nll & &
\times~
\left[ L(f_2,V_j)L(f_2,V_l)+R(f_2,V_j)R(f_2,V_l)\right].
\ea
We use the following conventions for the left- and right-handed couplings
between vector bosons and a fermion $f$:
\ba
\label{couplings}
 L(f,\gamma) &=& R(f,\gamma) =  \frac{e Q_f}{2},
\nll
 L(f,Z) &=& \frac{e}{4 s_W c_W}\left(2 I_3^f - 2 Q_fs^2_W \right),\ \ \
 R(f,Z) = \frac{e}{4 s_W c_W}\left(-2 Q_f s^2_W \right),
\nll
 L(q,g) &=& R(q,g) = \frac{1}{2} \sqrt{4\pi\alpha_s}.
\ea
The propagators are
\bq
D_V(s)=s-M_V^2+i\sqrt{s} \, \Gamma_V(s),
\label{propagator}
\eq
where $M_V$ and $\Gamma_V(s)=\sqrt{s}\Gamma_V/M_V$ are the mass and width
of the exchanged gauge boson ($M_\gamma=M_g=\Gamma_\gamma(s)=\Gamma_g(s)=0$).
Note that $\sigma_1(c,s,s_1,s_2)$ is of special interest because it describes
the production and decay of two off-shell $Z$-bosons, see fig.~1.
This is the numerically largest contribution, if the final fermions are not
both quark pairs.

For the interferences between different groups of diagrams ($k=4,5,6$),
we obtain
\ba
\frac{d^3\sigma_4(c,s,s_1,s_2)}{{\rm d}c{\rm d}s_1{\rm d}s_2} &=&
\frac{\sqrt{\lambda}}{\pi s^2}
 C_{233}(e,s;f_1,s_1;f_2,s_2){\cal T}_4 {\cal G}_{233}^c(c;s;s_1;s_2)~,\nll
\frac{d^3\sigma_5(c,s,s_1,s_2)}{{\rm d}c{\rm d}s_1{\rm d}s_2} &=&
\frac{\sqrt{\lambda}}{\pi s^2}
 C_{233}(f_1,s_1;e,s;f_2,s_2){\cal T}_6 {\cal G}_{233}^c(c;s_1;s;s_2)~,\nll
\frac{d^3\sigma_6(c,s,s_1,s_2)}{{\rm d}c{\rm d}s_1{\rm d}s_2} &=&
\frac{\sqrt{\lambda}}{\pi s^2}
 C_{233}(f_2,s_2;e,s;f_1,s_1){\cal T}_5 {\cal G}_{233}^c(c;s_2;s;s_1)~.
\ea
Again, the kinematical function ${\cal G}_{233}^c(c;s_2;s,s_1)$ is
given in the Appendix.
For $C_{233}$ we get
\ba
\label{c233}
C_{233}(e,s;f_1,s_1;f_2,s_2) &=&
 \frac{2 s s_1 s_2}{(6\pi^2)^2}
\, \Re e   \sum_{V_i,V_j,V_k,V_l=\gamma,Z}
\frac{1}{D_{V_i}(s)}\frac{1}{D_{V_j}(s_2)}
\frac{1}{D_{V_k}^*(s)}\frac{1}{D_{V_l}^*(s_1)}
\nll  &&\hspace{-2cm}
\times~
\left[ L(e,V_i)L(e,V_k)+R(e,V_i)R(e,V_k)\right]
\nll  &&\hspace{-2cm}
\times~
\left[ L(f_1,V_i)L(f_1,V_j)L(f_1,V_l)-R(f_1,V_i)R(f_1,V_j)R(f_1,V_l)\right]
\nll  &&\hspace{-2cm}
\times ~
\left[ L(f_2,V_j)L(f_2,V_k)L(f_2,V_l)-R(f_2,V_j)R(f_2,V_k)R(f_2,V_l)
\right].
\nll
\ea

The triple differential distributions are described by only one combination
of couplings for every interference,
exactly as the double differential case \cite{blr}.
However, the kinematical ${\cal G}$-functions are different.
They depend on one more
variable $c$ which singles out the electron-positron pair and makes the
${\cal G}$-functions less symmetric.

We are now prepared to write the formulae for various distributions,
\ba
\frac{{\rm d}\sigma}{{\rm d}c}&=&
\int_{s_{1}^-}^{s_{1}^+}{\rm d}s_{1}
\int_{s_2^-(s_1)}^{s_2^+(s_1)}{\rm d}s_2
\frac{{\rm d^3}\sigma(c,s,s_1,s_2)}{{\rm d}c{\rm d}s_1{\rm d}s_2},\nll
\frac{{\rm d}\sigma}{{\rm d}y_1}&=&
\int_{p_{1T}^-(y_1)}^{p_{1T}^+(y_1)}{\rm d}p_{1T}
\int_{s_2^-(p_{1T},y_1)}^{s_2^+(p_{1T},y_1)}{\rm d}s_2
\frac{{\rm d^3}\sigma(c,s,s_1,s_2)}{{\rm d}c{\rm d}s_1{\rm d}s_2},\nll
\frac{{\rm d}\sigma}{{\rm d}p_{1T}}&=&
\int_{y_{1}^-(p_{1T})}^{y_{1}^+(p_{1T})}{\rm d}y_{1}
\int_{s_2^-(p_{1T},y_1)}^{s_2^+(p_{1T},y_1)}{\rm d}s_2
\frac{{\rm d^3}\sigma(c,s,s_1,s_2)}{{\rm d}c{\rm d}s_1{\rm d}s_2}.
\ea
The kinematical borders of integration and the transformations from
$c$ and $s_1$ to $y_1$ and $p_{1T}$ are given in (\ref{vartrafo}) -
(\ref{borders2}).

The distributions
${\rm d}^3\sigma(c,s,s_1,s_2)/({\rm d}c{\rm d}s_1{\rm d}s_2)$
are derived for massless fermions. Hence, the formulae are not applicable
in the phase space regions
$s_1 \approx (m_1+m_2)^2,\ s_2 \approx (m_3+m_4)^2$.
However, the mass effects can be completely removed by a moderate cut on
$s_1$ and $s_2$. Such a cut should be applied to quark pairs in any case to
 remove non-perturbative QCD effects. To leading order, mass effects can be
taken into account as in ref. \cite{higgs}.

The numerically most important radiative corrections to the considered
process are due to initial state radiation of photons.
They may be taken into account with a convolution formula which is
well known from the $Z$ line shape~\cite{pittau,wwteup}.

\section{Summary}
In this paper, compact analytical formulae for triple differential
cross sections
${\rm d}^3\sigma(c,s,s_1,s_2)/({\rm d}c{\rm d}s_1{\rm d}s_2)$
of the process (\ref{reaction}) are presented. They make the semianalytic
approach more flexible allowing the calculation of many distributions
by at most two numerical integrations.
Further, it allows the inclusion of cuts on the integration variables
as summarized in the table:\vspace{0.3cm}\\
\begin{tabular}{|l|l|l|}\hline
observable & cuts possible on & No. of num. integrations \\ \hline
$\sigma$     & $s_1,s_2$                                        & 2 \\ \hline
${\rm d}\sigma/{\rm d}s_1$ & $s_2$                              & 1 \\ \hline
${\rm d}\sigma/{\rm d}c$ & $s_1,s_2$                            & 2 \\ \hline
${\rm d}\sigma/{\rm d}p_{1T}$ & $y_1,s_2$                       & 2 \\ \hline
${\rm d}\sigma/{\rm d}y_{1}$ & $p_{1T},s_2$                     & 2 \\ \hline
\end{tabular}
\vspace{0.3cm}\\
{\bf Table~1:} {\it Examples of cross sections and distributions
calculable by the semianalytic approach.}

In addition to the observables indicated in the table, double
differential distributions can also be calculated by less
numerical integrations. The triple differential distribution
${\rm d}^3\sigma/({\rm d}c{\rm d}s_1{\rm d}s_2)$ and the double
differential distribution ${\rm d}^2\sigma/({\rm d}s_1{\rm d}s_2)$
can be used as input for further investigations.

The formulae presented here, can easily be generalized to the exchange
of extra  heavy neutral gauge bosons by an extension of the
 summation in $C_{422}$ and $C_{233}$.
The generalization to longitudinal polarized beams is straightforward.

%
\appendix{\ Formulae for the kinematical functions}
%
We start with the kinematical function for {\tt crab} diagrams squared, see
fig.~1:
\bq
\label{g422c}
{\cal G}_{422}^{CC}(c;s;s_1,s_2) =
\frac{-1}{4}\left(\frac{1}{T_1}+\frac{1}{T_2}\right)
 \frac{s^2+(s_1+s_2)^2}{s-s_1-s_2}
-\frac{1}{4}\left(\frac{1}{T_1^2}+\frac{1}{T_2^2}\right)s_1s_2 -\frac{1}{2},
\eq
with
\bq
T_1=-\frac{1}{2}\left(s-s_1-s_2+c\sqrt{\lambda}\nobodyfrac\right),\ \
T_2=-\frac{1}{2}\left(s-s_1-s_2-c\sqrt{\lambda}\nobodyfrac\right).
\eq

The {\tt $f_1$-deer} diagrams squared lead to the function
\ba
\label{g422d}
{\cal G}_{422}^{DD}(c;s_1;s_2,s) &=&
\frac{3}{8}(1+c^2){\cal G}_{422}(s_1;s_2,s)
\nll
&&+\frac{1-3c^2}{\lambda}s_1(s+s_2)\frac{3}{4}
\left( 1 - 2{\cal L}(s_1;s_2,s)\frac{ss_2}{s_1-s_2-s} \right).
\ea
The logarithm ${\cal L}(s;s_1,s_2)$  is defined as
\bq
{\cal L}(s;s_1,s_2) =
\frac{1}{\sqrt{\lambda}} \, \ln \frac{s-s_1-s_2+\sqrt{\lambda}}
                                     {s-s_1-s_2-\sqrt{\lambda}}.
\label{L}
\eq
${\cal G}_{422}(s_1;s_2,s)$ is the corresponding kinematical function for the
double differential distribution $\frac{{\rm d}^2\sigma}
{{\rm d}s_1{\rm d}s_2}$
which we give here for sake of completeness \cite{blr}:
\bq
{\cal G}_{422}(s;s_1,s_2) =
 \frac{s^2+(s_1+s_2)^2}{s-s_1-s_2} {\cal L}(s;s_1,s_2) -2
\label{muta1}
\eq
The function of {\tt $f_2$-deers} squared is given by
${\cal G}_{422}^{DD}(c;s_2;s_1,s)$ which expresses the symmetry of the problem
according to the two final fermion pairs.

We can unify the two functions (\ref{g422c}) and (\ref{g422d}) in one function
 ${\cal G}_{422}^{c}(c;s_1;s_2,s)$ to be close to
 the old notation used for double differential distributions:
\ba
{\cal G}_{422}^{c}(c;s;s_1,s_2) = \left\{\begin{array}{ll}
{\cal G}_{422}^{CC}(c;s;s_1,s_2) & {\rm if}\ s>s_1\ {\rm and}\ s>s_2 \\
{\cal G}_{422}^{DD}(c;s;s_1,s_2) & {\rm else.}\end{array}\right.
\ea
As in the double differential case, ${\cal G}_{422}^c(c;s;s_1,s_2)$
is symmetric in the last two arguments.

The kinematical function of the interference between {\tt $f_1$-deers} and
{\tt $f_2$-deers} is
\ba
\label{g233dd}
 {\cal G}_{233}^{DD}(c;s;s_1,s_2) &=&
\frac{3}{8}(1+c^2) {\cal G}_{233}(s;s_1,s_2)\nll
&&- \frac{3}{\lambda^2}\frac{3}{8}(1-3c^2)s
\Biggl[ {\cal L}(s_1;s_2,s)2s_2(s_1-s_2)+(s-s_1-3s_2)\Biggr]\nll
&&\hspace{2cm}\times
\Biggl[ {\cal L}(s_2;s,s_1)2s_1(s_2-s_1)+(s-s_2-3s_1)\Biggr].
\ea
${\cal G}_{233}(s;s_1,s_2)$ is the kinematical function for
the double differential distribution \cite{blr}:
\ba
 {\cal G}_{233}(s;s_1,s_2) &=&
 \frac{3}{\lambda^2}\Biggl\{
 {\cal L}(s_2;s,s_1) {\cal L}(s_1;s_2,s)\nll &&\hspace{3cm}
          4s\left[ss_1(s-s_1)^2+ss_2(s-s_2)^2+s_1s_2(s_1-s_2)^2\right]
\nll
& & +~(s+s_1+s_2)\Biggl[
            {\cal L}(s_2;s,s_1)
          2s\left[(s-s_2)^2+s_1(s+s_2-2s_1)\right] \nll
& & \hspace{2.5cm} +~ {\cal L}(s_1;s_2,s)
          2s\left[(s-s_1)^2+s_2(s+s_1-2s_2)\right] \nll
& & \hspace{2.9cm} +~ 5s^2-4s(s_1+s_2)-(s_1-s_2)^2 \Biggr] \Biggr\}.
\label{muta2}
\ea

The interference between {\tt crabs} and {\tt $f_1$-deers} is given by:
\ba
\label{g233cd}
 {\cal G}_{233}^{CD}(c;s_2;s;s_1) &=&
 \frac{3}{\lambda^2}\Biggl\{
\frac{-1}{4}\left(\frac{1}{T_1}+\frac{1}{T_2}\right) {\cal L}(s_1;s_2,s)
          \nll &&\hspace{1cm}
\times 4s_2 \left[ss_1(s-s_1)^2+ss_2(s-s_2)^2+s_1s_2(s_1-s_2)^2\right]
\nll
& & \hspace{-0.5cm}
+~(s+s_1+s_2)\Biggl[ \frac{-1}{4}\left(\frac{1}{T_1}+\frac{1}{T_2}\right)
          2s_2\left[(s-s_2)^2+s_1(s+s_2-2s_1)\right] \nll
& & \hspace{2.5cm}
         +~     {\cal L}(s_1;s_2,s)
          s_2\left[(s_1-s_2)^2+s(s_1+s_2-2s)\right] \nll
& & \hspace{2.9cm} +~
\frac{1}{2}[ 5s_2^2-4s_2(s+s_1)-(s-s_1)^2 ]\Biggr] \Biggr\}.
\ea
The interference between {\tt crabs} and {\tt $f_2$-deers} is
expressed by ${\cal G}_{233}^{CD}(c;s_1;s;s_2)$.
In contrast to the double differential case, ${\cal G}_{233}^{CD}(c;s_2;s;s_1)$
 is not symmetric under the exchange of the last two arguments.
Again we write the functions (\ref{g233dd}) and (\ref{g233cd})
formally as one:
\ba
{\cal G}_{233}^c(c;s;s_1;s_2) = \left\{\begin{array}{ll}
{\cal G}_{233}^{DD}(c;s;s_1,s_2) & {\rm if}\ s>s_1\ {\rm and}\ s>s_2 \\
{\cal G}_{233}^{CD}(c;s;s_1;s_2) & {\rm else.}\end{array}\right.
\ea

All kinematical functions are finite and independent of $c$
for $\lambda\rightarrow 0$ (unfortunately, in ref. \cite{blr} is a typing
error in the formula for the limit of ${\cal G}_{233}(s;s_1;s_2)$):
\ba
{\cal G}_{422}^c(c;s;s_1,s_2) &\rightarrow & \frac{s(s_1+s_2)}{2s_1s_2}\nll
{\cal G}_{233}^{c}(c;s;s_1;s_2)  &\rightarrow &
\frac{s_1+s_2-5s}{(s_1-s_2-s)(s_2-s-s_1)}
\ea

Finally, we remark that the above functions give those of the double
differential distributions, if integrated over $c$:
\ba
\int_{-1}^1{\rm d}c\ {\cal G}_{422}^{c}(c;s;s_1,s_2)={\cal G}_{422}(s;s_1,s_2),
\nll
\int_{-1}^1{\rm d}c\ {\cal G}_{233}^{c}(c;s;s_1;s_2)={\cal G}_{233}(s;s_1,s_2).
\ea
\end{document}